# A demographic scaling model for estimating the total number of COVID-19 infections

Christina Bohk-Ewald[a,b], Christian Dudel[b], Mikko Myrskylä[a,b]

[a] Center for Social Data Science, University of Helsinki, Unioninkatu 35, 00170 Helsinki, Finland
[b] Max Planck Institute for Demographic Research, Konrad-Zuse-Str. 1, 18057 Rostock, Germany

**Abstract**

Understanding how widely COVID-19 has spread is critical for examining the pandemic's progression. Despite efforts to carefully monitor the pandemic, the number of confirmed cases may underestimate the total number of infections. We introduce a demographic scaling model to estimate COVID-19 infections using an broadly applicable approach that is based on minimal data requirements: COVID-19 related deaths, infection fatality rates (IFRs), and life tables. As many countries lack reliable estimates of age-specific IFRs, we scale IFRs between countries using remaining life expectancy as a marker to account for differences in age structures, health conditions, and medical services. Across 10 countries with most COVID-19 deaths as of May 13, 2020, the number of infections is estimated to be four [95% prediction interval: 2-11] times higher than the number of confirmed cases. Cross-country variation is high. The estimated number of infections is 1.4 million (six times the number of confirmed cases) for Italy; 3.1 million (2.2 times the number of confirmed cases) for the U.S.; and 1.8 times the number of confirmed cases for Germany, where testing has been comparatively extensive. Our prevalence estimates, however, are markedly lower than most others based on local seroprevalence studies. We introduce formulas for quantifying the bias that is required in our data on deaths in order to reproduce estimates published elsewhere. This bias analysis shows that either COVID-19 deaths are severely underestimated, by a factor of two or more; or alternatively, the seroprevalence based results are overestimates and not representative for the total population.

**Introduction**

The total number of COVID-19 infections is a key indicator for understanding the spread of the pandemic. Despite its central importance in policy decisions, this indicator is largely unknown. The number of confirmed COVID-19 cases may underestimate the total number, and the few seroprevalence studies for COVID-19 have relied on non-representative samples.[1-4] Moreover, estimates that are based on complex statistical methods and simulation approaches[5,6] provide important information, but have high data demands, and are not straightforward to implement.

We introduce a demographic scaling approach to estimate the number of COVID-19 infections and their fraction in a population. This indirect approach can be applied in many contexts, as it requires only a small amount of input data: namely, the number of COVID-19-related deaths for the population of interest; and the age-specific infection fatality rates (IFR; deaths over infections) from a reference population, scaled to match the target population based on life tables.

Scaling IFRs is not only a key and novel feature of the introduced approach, it is necessary, as age-specific IFRs are not available for many countries. The scaling step allows us to transfer the best IFR estimates available globally from a reference population to a target population. The IFRs are scaled to indirectly adjust for underlying differences in age structures, health conditions, and health care services. The scaling makes use of the demographic concept of remaining life expectancy, which is sometimes called thanatological age. It maps IFRs between a reference and a target country so that the people in a given age group in the target population are assigned the same IFR as the people in another age group in the reference population if they have the same number of remaining life years.

We apply the demographic scaling model to estimate the total numbers of COVID-19 infections as well as their population fractions in 10 countries that have the most reported COVID-19 deaths as of May 13, 2020. The model estimates suggest that the COVID-19 pandemic is much more spread in each of those countries than the number of confirmed cases suggests. We also find that the population fractions have increased exponentially with time, reaching levels as high as 3.1% [95% prediction interval: 2.6%-7.7%] in Belgium, but that they are far from reaching levels required for herd immunity.

The demographic scaling model can also be used to validate the quality of infection prevalence estimates based on other approaches. We do this by developing formulas for quantifying the bias that is required in our data on deaths or IFRs in order to reproduce infection estimates that are based on local seroprevalence studies or simulation models published elsewhere. This bias quantification can be used both for validating our model results, and as a benchmark for assessing the quality of infection estimates from other sources. Our model estimates of the COVID-19 prevalence are considerably lower than the estimates of recent local seroprevalence studies for the U.S., Italy, and Germany. This discrepancy can result either from the input data (deaths or IFRs) in our approach being wrong, or the local seroprevalence studies not being representative. Using our model to assess the bias required to reproduce the results of these local seroprevalence studies indicates either a very high undercount of deaths, e.g., fewer than two in 100 deaths for Germany, which is perhaps unlikely; or that the local infection estimates were not representative for the total population on the national level.

Readers have access to the R source code and information on the data at https://github.com/christina-bohk-ewald/demographic-scaling-model.

**The empirical relationship between confirmed cases and deaths from COVID-19**

Figure 1 shows the relationship between confirmed cases and deaths from COVID-19 for the 10 countries that have reported the most COVID-19 deaths as of May 13, 2020. The countries fan out between Belgium and Germany, and their crude case fatality rates (CFR; deaths over confirmed cases) range between 4.5% and 16.4%. Up to this date, Belgium, France, the U.K., Italy, the Netherlands, and Spain had experienced more COVID-19 deaths per confirmed cases than Brazil, the U.S., Iran, or Germany. This variation in CFRs could be driven by several factors, some of which are "real" differences arising from differences in the age-specific mortality risks among the infected.[7,8,9] However, the variation in CFRs may also reflect other factors, such as differences in testing intensity[10,11] and test specificity;[11,12] variation in terms of the age structure of the number of confirmed cases;[8] the stage of progress of the COVID-19 outbreak;[1] and differences in the classification of deaths. Roser and colleagues[13] discussed most of these issues.

FIGURE 1

For our purposes, a key question is whether the numerator of the CFR, or the number of deaths, is more or less accurate than the denominator, or the number of confirmed cases. The number of confirmed cases may strongly underestimate the total number of COVID-19 cases.[11,14] For example, cases with mild symptoms or asymptomatic cases may go undetected; test coverage may be poor, and may focus on specific sub-populations only, or on tracing back only to people with proven contact to confirmed COVID-19 cases; and false negatives may outnumber false positives.[11,12]

On the other hand, the number of COVID-19 deaths may be under- or overestimated. Reporting delays and inconsistent practices for defining COVID-19 deaths are among the key sources of error. For example, in some contexts, all deceased individuals who have had COVID-19 may be counted, while in other contexts, only deceased individuals for whom COVID-19 was the primary cause of death, or only deceased individuals who had been hospitalized for treatment of COVID-19, may be counted. Testing practices and coverage may also be insufficient to detect all of the individuals who have died from COVID-19.[13] This could lead to COVID-19 deaths being undercounted. Studies that analyze excess mortality would be helpful for quantifying this bias.[15] We argue, however, that the numbers of COVID-19 deaths are more reliable than the numbers of confirmed COVID-19 cases, and thus use COVID-19 deaths as the core empirical input in our estimation approach. However, since the coverage of COVID-19 deaths is difficult to assess, we develop formulas for assessing the potential effects of the over- and the underreporting of COVID-19 deaths on our results.

**Demographic scaling approach to estimate COVID-19 infections**

We introduce a demographic scaling approach to estimate the total number of COVID-19 infections. This approach is built on the assumption that COVID-19 deaths are fairly accurately recorded, and that IFRs borrowed from a reference country reflect the true IFRs of the target country after appropriate scaling. Each of these assumptions can be criticized, and we do so below in the Discussion.

We start with the basic identity that represents the age-specific number of infected:

(1) $I_x = P_x \cdot \lambda_x,$

where $I$ is the unknown number of infected, P is the known population size, $\lambda$ is the unknown fraction of the infected population, and x denotes the age group. We estimate $\lambda_x$ by using the equation $D_x = IFR_x \cdot P_x \cdot \lambda_x$, where $D$ is the number of deaths and $IFR$ is the infection fatality rate. We rearrange the equation to get $\lambda_x = D_x / [IFR_x \cdot P_x]$, and estimate the total number of infected by

(2) $I = \sum_x P_x \cdot \lambda_x$

Replacing $\lambda_x$ with its definition yields

(3) $I = \sum_x D_x / IFR_x$

The key challenge is to arrive at credible estimates of $IFR_x$ and $D_x$.

The simplest way to obtain estimates of $IFR_x$ is to borrow them from some source that appears to be valid for the country of interest. We call this approach the unadjusted approach, and consider it a potentially viable option only if the life table data needed for the scaling approach are not available. The unadjusted approach is risky, as IFRs that are valid in one context may not carry over to another context, even if they are age-specific. For example, the presence of underlying health conditions – such as cardiovascular diseases, diabetes, chronic respiratory diseases, hypertension, and cancer[16] – increase the risk of death of a given COVID-19 infection, and the distribution of these conditions varies. In addition, the ability of health care systems to treat illnesses effectively may differ across countries.

A superior approach is to adjust the IFRs taken from one context to reflect the specific age structure, health status, and health care system of the target country. To control for such cross-country differences, we map IFRs between two countries based on their remaining lifetime (or thanatological age), denoted by $e_x$. More specifically, we assign the same infection fatality rate (IFR) to people of two countries who have, on average, the same number of life years left ($e_x$):

$$IFR_{e_x}^{COI} = IFR_{e_x}^{RC},$$

where COI denotes the country of interest and RC denotes the reference country. For example, if 70-year-olds in a reference country have, on average, the same number of life years left as 75-year-olds in a country of interest, the infection fatality rate of the 70-year-olds in the reference country is used for the 75-year-olds in the country of interest.

Mapping infection fatality rates based on remaining lifetime allows us to adjust for cross-country differences in age and health structures, as well as in medical care. That is because remaining life time is a function of underlying health conditions and a health care system's effectiveness in curing them.[17] The fewer underlying health conditions people have and the more effective medical care is in treating them, the healthier people are and the more life years they have left.

Another data challenge is related to COVID-19 deaths, $D_x$. While these counts are available in total numbers for many countries worldwide,[18] they are often not available by age. To deal with this problem, we disaggregate the total deaths into age groups using a global average pattern over age that we estimated by analyzing all data on COVID-19 deaths by age provided by Dudel and colleagues[8] (*SI appendix* 3).

Deaths may be misreported. The impact on our results can be assessed by introducing the relative amount of under- or overreporting directly into formula (3),

(4) $I^B = B \sum_x D_x / IFR_x = B \cdot I^T,$

where $B$ captures under- and overreporting, assuming that misreporting affects all ages to the same extent; $I^T$ is the true number of infections; and $I^B$ is the number of infections observed with reporting bias. If $B$ is below one, then there is bias through underreporting; and if $B$ is above one, then there is bias through overreporting. Equation (4) shows that if there is an estimate of $B$, a biased estimate of infections can easily be adjusted to derive the true number, $I^T = I^B / B$. Equation (4) also allows us to calculate the misreporting of deaths that is required to "explain" an estimate of $I$ taken from another reference. For example, if our method provides an estimate $I^E$, while another method, such as a seroprevalence study, yields another estimate $I^S$, then to explain the difference between $I^E$ and $I^S$ through misreporting, while assuming that $I^S$ is the true value, $B$ has to equal $I^E / I^S$. Depending on the resulting value of $B$, the values of $I^E$ and $I^S$ might be considered inconsistent if $B$ is very high or low. This may indicate that the reporting of deaths is biased, or that the results of the seroprevalence study are not representative.

We use the scaling approach (1) to estimate the total number and prevalence of COVID-19 infections for the 10 countries that have reported the most deaths caused by COVID-19 as of May 13, 2020 and (2) to assess the validity of recent local seroprevalence studies for the U.S., Italy, and Germany. As input data, we use (1) 2019 population counts of the UN World Population Prospects;[19] (2) accumulated total COVID-19 deaths from Johns Hopkins University CSSE;[18] and (3) IFRs reported by Verity and colleagues[20] for Hubei. Verity and colleagues[20] conducted a Bayesian analysis, and provided credible intervals for their point estimates. We use these credible intervals to generate uncertainty estimates (prediction intervals) of the number of infected individuals. Both the population counts and the IFRs are provided by 10-year age groups, $0 - 9$, $10 - 19$, ..., $80 +$. We disaggregate COVID-19 total deaths into the same 10-year age groups using the global average pattern over age that we estimated based on the data provided by Dudel and colleagues.[8] Details about the model and additional findings based on, e.g., unadjusted IFRs, adjusted IFRs from a European reference country as reported by Salje and colleagues,[21] and average time to death from COVID-19 are given in *SI appendix* 1 through 6.

**Results**

Figure 2 shows the numbers of confirmed and estimated COVID-19 infections as of May 13, 2020. Our estimates suggest that across the countries, the total number of infections is approximately four times higher than the number of confirmed cases. For example, for the U.S., which has 1.4 million confirmed cases, we estimate that the total number of infections might range from approximately 1.4 million to 6.7 million, with a central estimate of 3.1 million infections. For a large number of countries, the central estimate for the number of infections is more than five times higher than the number of confirmed cases. For example, for Italy, we estimate that there are approximately 1.4 million infections, whereas the total number of confirmed cases is just 222k. Germany, where testing has been comparatively extensive, stands out in this context, as our results suggest that the number of infections in Germany is only 1.8 times higher than the number of confirmed cases (305k versus 174k).

FIGURE 2

Our lower bound estimates suggest that across the countries in our sample, the total number of infections is approximately twice as high as the number of confirmed cases. For example, for Italy, the lower bound is 590k, or 2.7 times the number of confirmed cases (222k). For France, our lower bound estimate is 536k infections, which is 3.1 times higher than the number of confirmed cases (176k). In Germany, our lower bound estimate suggest that the total number of infections might be lower, at 138k, than the number of confirmed cases (174k). Note, however, that the confirmed cases within 95% prediction intervals do not indicate model error, but rather effective testing in a country, as many people who are or have been infected with COVID-19 are detected.

The upper bound estimates for the total number of confirmed cases suggest that the total number of infections may be more than 11 times higher than the number of confirmed cases in some countries. For the U.S., our results suggest that there is an upper bound of 6.7 million, which is almost five times higher than the number of confirmed cases. In Belgium, Italy, and France, our upper bound estimates are 892k, 3.7 million, and 3.3 million, respectively, or more than 16 to 18 times higher than the numbers of confirmed cases. In addition, for Spain, the U.K., and the Netherlands, our upper bound estimates are more than 13 to 16 times higher than the number of confirmed cases. Iran, Germany, and Brazil are the only countries where the upper bound is less than five times higher than the number of confirmed cases.

Figure 3 shows the COVID-19 prevalence estimates over time. For each date shown, the numbers may underestimate the fractions of the population who are infected, since we do not account for the time lag between infection and death. However, accounting for the time lag is not expected to strongly change the shape of the curves (as shown in *SI appendix* 6). Based on these estimates, as of May 13, 2020, we find the central prevalence to be on average 1.5%. It ranges from 3.1% in Belgium; to between 1.9% and 2.7% in Spain, Italy, France, and the U.K.; to approximately 1.3% in the Netherlands; to 0.9% in the U.S. and to 0.4% or less in Germany, Iran, and Brazil. The uncertainty bounds are wide and range on average between 0.7% and 3.8%. The lower bound includes values as low as 0.01%, 0.1%, and 0.2% for Brazil, Iran, and Germany, respectively. The upper bound includes values as high as 7.8%, 7.7%, and 6.1% for Spain, Belgium, and Italy, respectively.

FIGURE 3

Considering both the total numbers and prevalence of COVID-19 infections provides two complementary perspectives on the progress of the pandemic within and between countries. While we estimate for the U.S. to have the largest numbers of infections across our sample of ten countries, as of May 13, 2020, we also estimate for the U.S. to have a comparatively small prevalence of individuals who are infected with COVID-19. We find an opposite pattern for Belgium, a country that has almost 30 times fewer inhabitants than the U.S.. We estimate for Belgium to have comparatively small total numbers of infections, but comparatively high infection prevalence. In fact, we estimate the population fractions of COVID-19 infections to be largest in Belgium since April 22, 2020, so that they are even larger than in Spain and Italy who are estimated to have had the largest COVID-19 prevalence since April 2, 2020 and February 24, 2020, respectively, as COVID-19 has started to spread earlier in those two countries.

Compared to the results of recent local seroprevalence studies for the U.S., Italy, and Germany, our COVID-19 prevalence estimates are much lower for those countries. We assess the quality of these seroprevalence-based infection estimates with our demographic scaling model by showing how many more COVID-19 deaths would have been required to match them. For example, Bendavid and colleagues[3] reported in the middle of April 2020 a seroprevalence between 1.1% and 5.7% for Santa Clara County in the U.S., compared to our central estimate of 0.4% for the U.S. as a whole at the same time (April 17, 2020). Assuming the seroprevalence estimate of 1.1% is correct, we would need to have only one in three COVID-19-related deaths registered. Bendavid and colleagues[3] estimated a seroprevalence of 10% for the city of Robbio in Italy, and a seroprevalence of 14% for the German municipality of Gangelt. To be compatible with our central prevalence estimates of 1.7% in Italy and 0.2% in Germany, only one in five COVID-19-related deaths would have to be recorded in Italy, and fewer than two in 100 deaths would have to be recorded in Germany. However, when working with the upper bound of our prevalence estimates, only one in two COVID-19-related deaths would have to be missed for the U.S. and Italy, which is possible; whereas for Germany, the number is still unrealistically high. Either way, the estimates based on local, non-representative seroprevalence studies appear to be higher than our prevalence estimates, which may suggest that they are not representative of the total population.

**Discussion**

The total number of infections is among the key unknowns of the COVID-19 pandemic. Several studies have provided estimates of the number and prevalence of COVID-19 infections, based either on local seroprevalence measurement[3,4] or complex projection models.[5,6] We have developed a demographic scaling model to estimate the number and prevalence of infections based on modest data requirements. Our results suggest that across the 10 countries with most COVID-19 deaths as of May 13, 2020, the total number of infected individuals is approximately four times higher than the number of confirmed cases. However, the level of uncertainty about these findings is high, as the lower and upper bound of the 95% prediction interval suggest that, on average, there are twice and 11 times as many infections as confirmed cases, respectively. Country-specific variation is also high. Our estimates for Italy suggest that the total number of infections is approximately 1.4 million, or more than six times higher than the country-specific number of confirmed cases. For the U.S., our estimate of 3.1 million infections is more than twice as high as the number of confirmed cases, and the upper bound of 6.7 million is almost five times higher than the number of confirmed cases. For Germany, where testing has been comparatively extensive, we estimate that the total number of infections is only 1.8 times (upper bound: slightly more than four times) higher than the number of confirmed cases.

Considering the urgent need for population-based seroprevalence studies in order to assess the progress of the COVID-19 pandemic and to determine optimal time points for (de-)escalating control measures, it is also critically important to have at hand or to introduce evaluation tools for them. Our demographic scaling approach is also a suitable tool to validate infection prevalence estimates that have been published based on seroprevalence studies. It can be used to analyze (1) how many more or fewer deaths or (2) how much higher or lower infection fatality rates would need to be for our results to match those from seroprevalence studies. A comparison of our model estimates for the U.S., Italy, and Germany with results from recent local seroprevalence studies shows that the latter usually provide rather high prevalence estimates that are unlikely to apply to the total population, particularly in the case of Germany. Local seroprevalence studies may be biased due to various reasons, such as false negative and false positive test results, and population samples that are not nationally representative.[22]

Our model estimates of COVID-19 infections build on two key assumptions: (1) total numbers of deaths from COVID-19 are accurately recorded, and (2) the scaled infection fatality rates from China reported by Verity and colleagues[20] (or from another source) can be applied to other countries. At best, these assumptions only partially hold.

Our first key assumption implies that COVID-19 deaths are fairly accurately recorded. However, COVID-19 deaths may be underestimated, particularly in regions that are heavily affected by the pandemic.[23] Reporting delays that can be as long as several days may result in the underestimation of COVID-19 deaths in all regions. Inconsistent and changing practices for defining COVID-19 deaths and poor test coverage may also influence the accuracy of death counts.[13] However, as we argue that the numbers of COVID-19 deaths are more reliable than the numbers of confirmed COVID-19 cases, we selected death counts as the core empirical input in our estimation approach. If the numbers of total deaths were too small, the estimated numbers of infections would be biased downward, and vice versa. However, if the deaths were misreported and the amount of bias caused by this misreporting was known, our approach could easily incorporate this information. Studies that analyze excess mortality are urgently needed,[15] as they may help to gauge the amount of bias in COVID-19 death counts.

Our second key assumption implies that infection fatality rates from a reference country become applicable in a country of interest through scaling based on remaining life expectancy, and that these life years left are a useful marker to account for overall cross-country differences in age structure, health conditions, and medical services. Although scaling infection fatality rates between a reference and a target country increases the applicability of this estimation approach, such borrowing strategies do not fully reflect country-specific trends. It is also important to note that using the infection fatality rates of epidemiological studies (based on nasopharyngeal swabs or, even better, population-representative serological studies) would be preferable to relying on estimates of other models, as it would enable us to avoid the circling effects between different modeling approaches. However, infection fatality rates are lacking for many countries so that we make use of those derived from a Bayesian model for Hubei, China, by Verity and colleagues,[20] which have also been adopted in the context of the U.K. and the U.S.. *SI appendix* 5 shows how estimations of COVID-19 infections increase when they are based on scaling French infection fatality rates from a Bayesian model reported by Salje and colleagues.[21] Infection fatality rates can be biased due to, e.g., model misspecification and misreported input data.

In addition to relying on the two key assumptions regarding death counts and infection fatality rates, we ignore the time lag between infection (or reporting date) and death, which may be several weeks.[24-26] One option for adjusting for the time lag between the onset of infection and death is to compare estimated infections with confirmed cases 18 days ago, as Zhou and colleagues[25] found this to be the average duration until death from COVID-19. Accounting for this time lag would lead to higher estimates of infections (*SI appendix* 6). However, finding the correct counterpart of confirmed cases is not straightforward, as reporting dates and data about time to death vary.[24,25]

Developing simple and fast approaches to estimating the total number and prevalence of infections is important in situations in which the detailed data needed for precise estimation are not available and population-representative seroprevalence studies are lacking. Our demographic scaling model can be implemented broadly, and provides useful information about the magnitude of the unknown number of infections and their population fractions in countries across the globe. It is also a suitable tool to validate infection estimates based on model simulations and seroprevalence studies. Our model outcomes can be used in decision-making, and as input in more advanced models.[1,5,6,27-29] Moreover, as the information about the key parameters of the estimation approach – deaths and infection fatality rates – improves, the approach will produce increasingly accurate results.

**Acknowledgments**

We are grateful for the input of Alyson van Raalte and Enrique Acosta from the Max Planck Institute for Demographic Research, Germany, on an early draft of our manuscript.

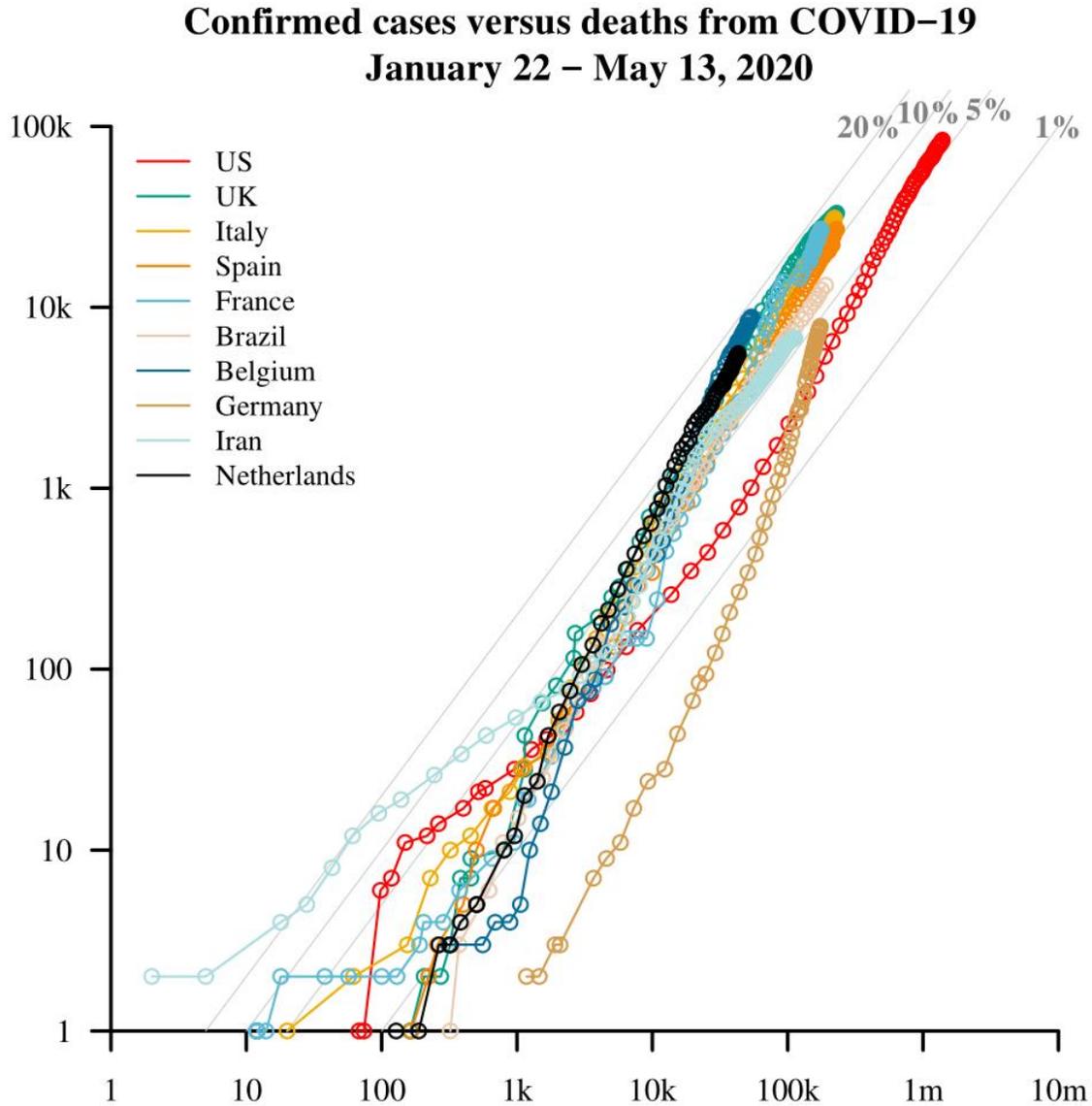

**Figure 1.** Confirmed cases on the horizontal axis versus deaths attributable to COVID-19 on the vertical axis, on the logarithmic scale, from January 22 to May 13, 2020. Different levels of the case fatality rate, in %, are highlighted with gray lines and text. Shown are the 10 countries that have the largest numbers of reported deaths from COVID-19 as of May 13, 2020. Data: Johns Hopkins University CSSE.[18] Own calculations.

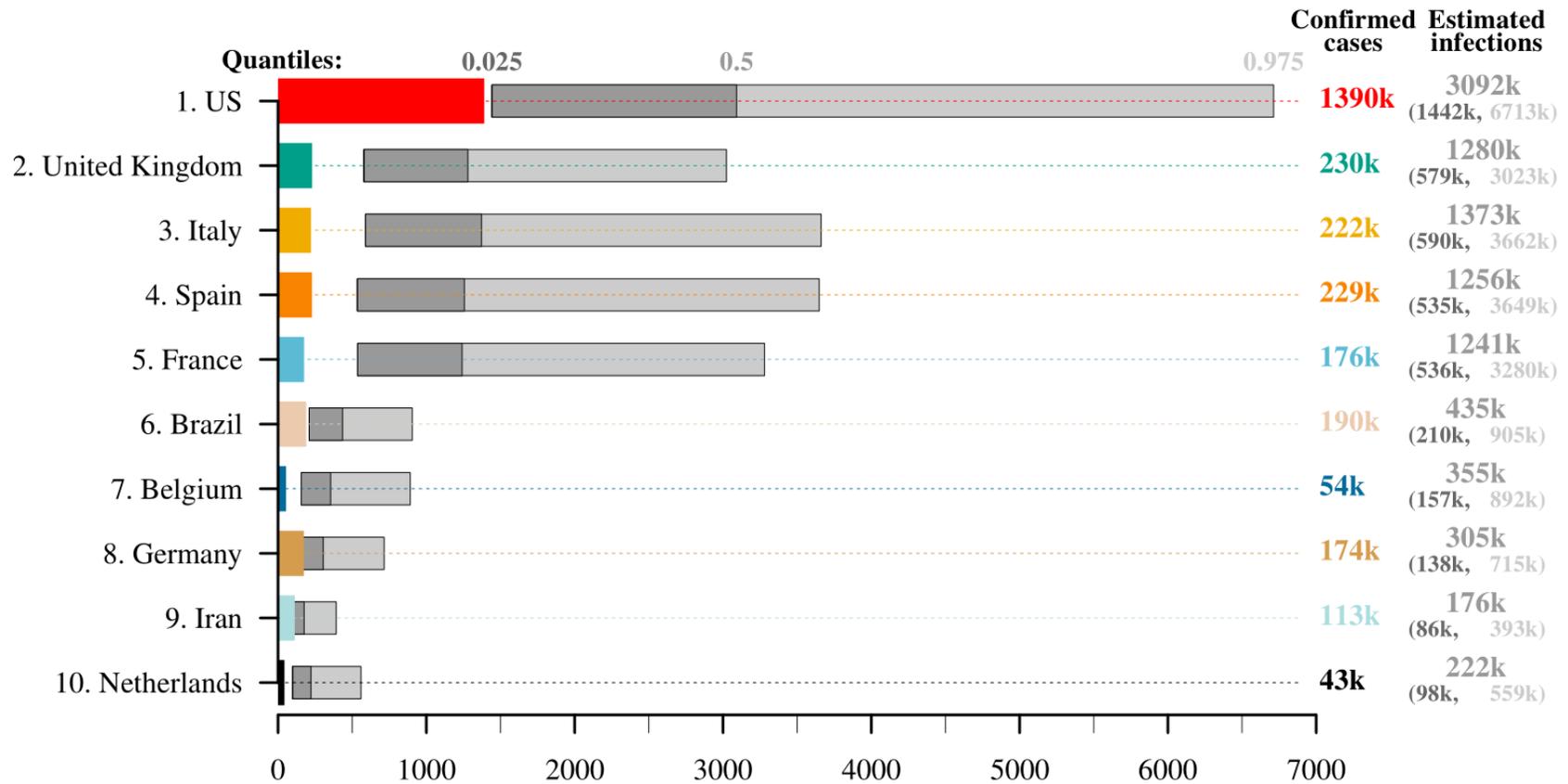

**Figure 2.** Confirmed cases and estimated total number of COVID-19 infections, from January 22 to May 13, 2020, for the 10 countries that have the largest numbers of reported deaths from COVID-19 as of May 13, 2020. Own calculations using data from Verity and colleagues,[20] UN World Population Prospects,[19] and Johns Hopkins University CSSE.[18]

**Fraction of people probably infected with COVID−19, January 22 − May 13, 2020**

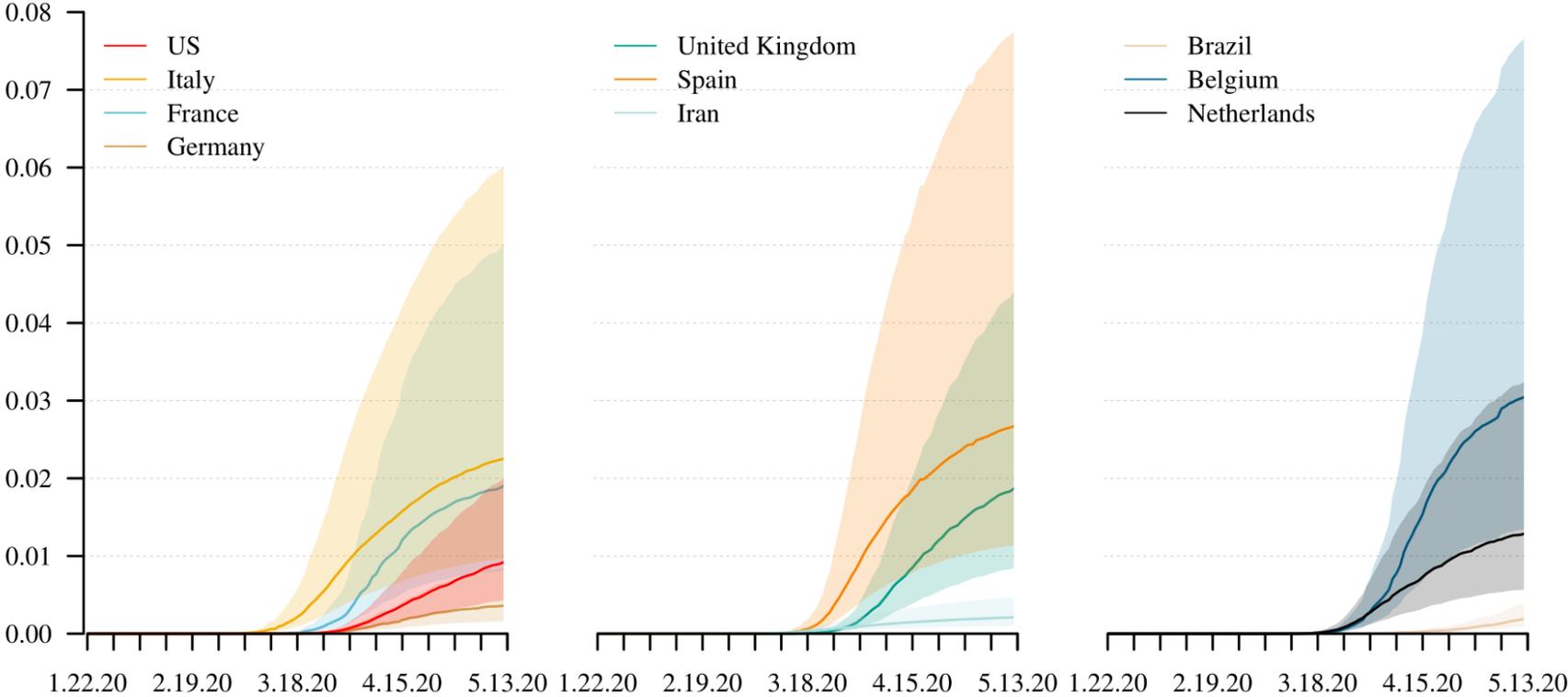

**Figure 3.** Estimated population share of COVID-19 infections, from January 22 to May 13, 2020, for the 10 countries that have the largest numbers of reported deaths from COVID-19 as of May 13, 2020. Own calculations using estimates of Verity and colleagues,[20] UN World Population Prospects,[19] and Johns Hopkins University CSSE.

**SUPPLEMENTARY INFORMATION**

**SI appendix 1**

**COVID-19 infection fatality rates**

To estimate COVID-19 infections with the introduced demographic scaling model for the 10 countries with most COVID-19 deaths as of May 13, 2020, we map the infection fatality rates of Hubei, China, as published by Verity and colleagues.[1] For the sake of convenience and transparency, we list these infection fatality rates (the mode as well as the lower and upper bound of the 95% credible interval) in Table S1.

**Table S1: Infection fatality rates observed in Hubei, China**

| | Infection fatality rates, Hubei, China | | |
| Age group | Mode | Lower 95% | Upper 95% |
| --- | --- | --- | --- |
| 0-9 | 0.000016 | 0.00000185 | 0.000249 |
| 10-19 | 0.00007 | 0.000015 | 0.0005 |
| 20-29 | 0.00031 | 0.00014 | 0.00092 |
| 30-39 | 0.00084 | 0.00041 | 0.00185 |
| 40-49 | 0.0016 | 0.00076 | 0.0032 |
| 50-59 | 0.006 | 0.0034 | 0.013 |
| 60-69 | 0.019 | 0.011 | 0.039 |
| 70-79 | 0.043 | 0.025 | 0.084 |
| 80+ | 0.078 | 0.038 | 0.133 |

Infection fatality rates (mode and lower and upper bound of 95% credible interval) by 10-year age groups observed in Hubei, China. Data source: Verity and colleagues.[1]

**SI appendix 2**

**Estimate COVID-19 infections based on scaling infection fatality rates between two countries via thanatological age**

It takes four steps to estimate COVID-19 infections for a country of interest with the introduced demographic scaling model, which maps the infection fatality rates of Hubei, China, onto a country of interest via the thanatological age.

1. Ungroup the reference country's infection fatality rates $IFR_x$ from 10-year age groups into single years of age using a cubic smoothing spline via the R-function *smooth.spline*.

2. Ungroup the remaining life years ($e_x$), taken from abridged life tables of the UN World Population Prospects,[2] for both the reference country and the country of interest.

3. Map the ungrouped infection fatality rates of the reference country onto the country of interest via thanatological age. The mapped infection fatality rates for the 10 countries with the most COVID-19 deaths as of May 13, 2020, are shown in Figure 1.

4. Calculate the number of COVID-19 infections ($I$) based on equation $I = \sum_x D_x / IFR_x$.

Figure S1 displays the mapped IFR$_x$ for the 10 countries with the most deaths attributable to COVID-19 as of May 13, 2020. Tables S2 and S3 list the corresponding IFR$_x$ by 10-year age groups, as well as the crude IFR$_x$ for each of those countries, based on central estimates and the upper bound of the 95% prediction interval.

## Scaled infection fatality rates based on remaining life expectancy
## Reference country: Hubei, China

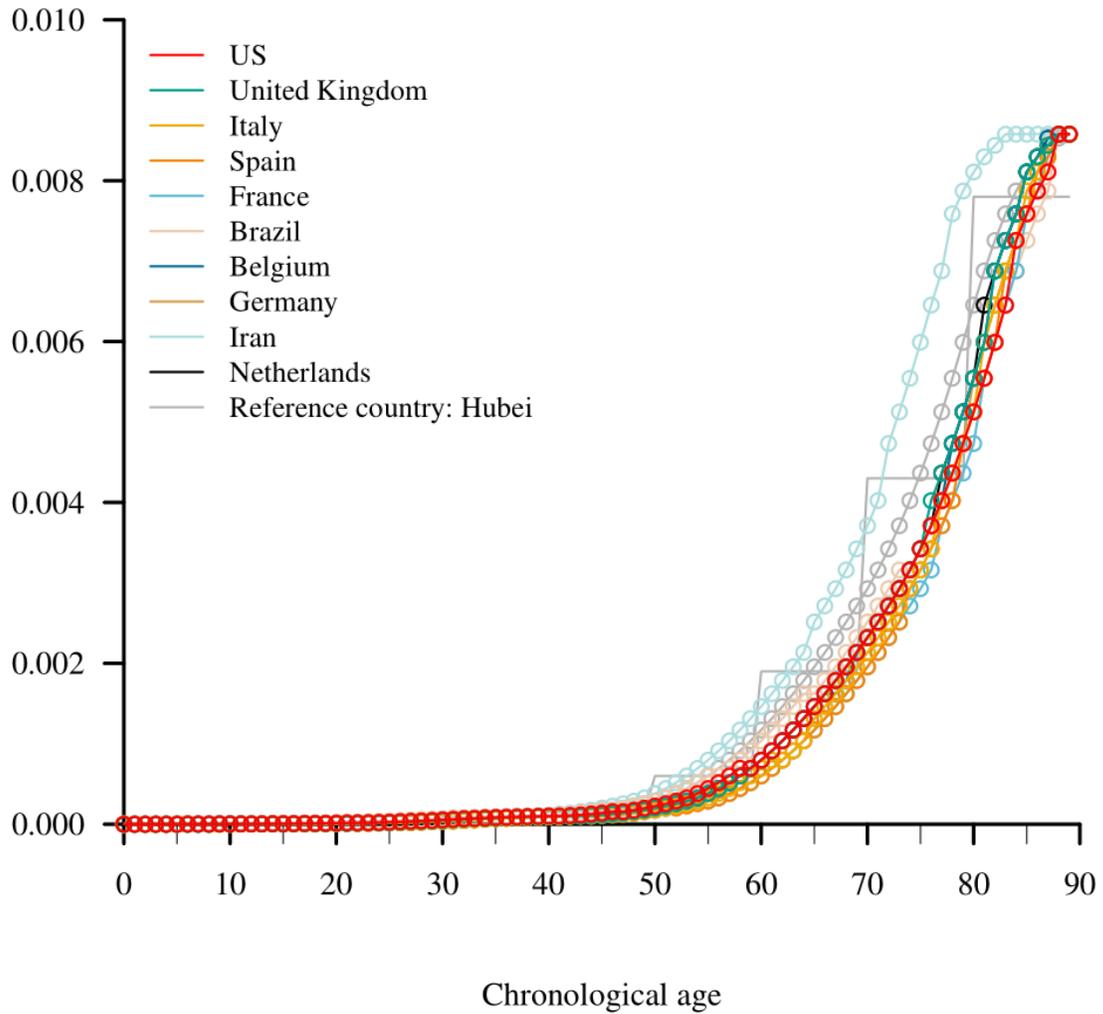

**Fig. S1.** Shown are the 10 countries that have the largest numbers of reported deaths from COVID-19 as of May 13, 2020. Own calculations using data from Verity and colleagues[1] and abridged life tables of UN World Population Prospects.[2]

**Table S2: Scaled central estimates of IFR for the 10 countries with most COVID-19 deaths as of May 13, 2020**

| | Age-specific Infection Fatality Rates | | | | | | | | | Crude IFR |
|---|---|---|---|---|---|---|---|---|---|---|
| | 0-9 | 10-19 | 20-29 | 30-39 | 40-49 | 50-59 | 60-69 | 70-79 | 80+ | All ages |
| US | 0.000016 | 0.00007 | 0.00028 | 0.00079 | 0.00138 | 0.00442 | 0.01420 | 0.03389 | 0.071095 | 0.027208 |
| Italy | 0.000010 | 0.000037 | 0.00017 | 0.00063 | 0.00113 | 0.00341 | 0.01238 | 0.03232 | 0.073279 | 0.022662 |
| Spain | 0.000010 | 0.000033 | 0.00017 | 0.00063 | 0.00113 | 0.00301 | 0.01140 | 0.03090 | 0.071946 | 0.021577 |
| France | 0.000010 | 0.000038 | 0.00017 | 0.00063 | 0.00113 | 0.00310 | 0.01140 | 0.02983 | 0.070512 | 0.021782 |
| UK | 0.000011 | 0.000051 | 0.00023 | 0.00074 | 0.00123 | 0.00393 | 0.01420 | 0.03531 | 0.075259 | 0.025927 |
| Belgium | 0.000011 | 0.000044 | 0.0002 | 0.00069 | 0.00122 | 0.00393 | 0.01386 | 0.03465 | 0.075353 | 0.024944 |
| Iran | 0.000021 | 0.000092 | 0.00036 | 0.00091 | 0.00201 | 0.00787 | 0.02370 | 0.05792 | 0.084891 | 0.03859 |
| Hubei | 0.000018 | 0.000081 | 0.00036 | 0.00088 | 0.00178 | 0.00689 | 0.02076 | 0.04653 | 0.080126 | 0.034729 |
| Germany | 0.000012 | 0.000052 | 0.00023 | 0.00074 | 0.00132 | 0.00395 | 0.01420 | 0.03389 | 0.072849 | 0.025759 |
| NL | 0.000011 | 0.000044 | 0.0002 | 0.00069 | 0.00122 | 0.00393 | 0.01420 | 0.0350 | 0.075814 | 0.025072 |

Own calculations based on the age-specific modal IFR of Verity and colleagues[1] and abridged life tables of the UN World Population Prospects.[2]

**Table S3: Scaled upper 95% estimates of IFR for the 10 countries with most COVID-19 deaths as of May 13, 2020**

| | Age-specific Infection Fatality Rates | | | | | | | | | Crude IFR |
|---|---|---|---|---|---|---|---|---|---|---|
| | 0-9 | 10-19 | 20-29 | 30-39 | 40-49 | 50-59 | 60-69 | 70-79 | 80+ | All ages |
| **US** | 0.000002 | 0.000015 | 0.00012 | 0.00039 | 0.0006 | 0.0025 | 0.0080 | 0.0204 | 0.0357 | 0.0125 |
| **Italy** | 0.000002 | 0.000004 | 0.00006 | 0.00032 | 0.0005 | 0.0019 | 0.0070 | 0.0194 | 0.0364 | 0.0085 |
| **Spain** | 0.000002 | 0.000003 | 0.00006 | 0.00032 | 0.0005 | 0.0017 | 0.0064 | 0.0185 | 0.0360 | 0.0074 |
| **France** | 0.000002 | 0.000004 | 0.00006 | 0.00032 | 0.0005 | 0.0017 | 0.0064 | 0.0179 | 0.0355 | 0.0082 |
| **UK** | 0.000002 | 0.000008 | 0.00010 | 0.00037 | 0.0005 | 0.0022 | 0.0080 | 0.0210 | 0.0371 | 0.0110 |
| **Belgium** | 0.000002 | 0.000006 | 0.00008 | 0.00035 | 0.0005 | 0.0022 | 0.0078 | 0.0207 | 0.0371 | 0.0099 |
| **Iran** | 0.000002 | 0.000024 | 0.00017 | 0.00043 | 0.0010 | 0.0044 | 0.0141 | 0.0310 | 0.0403 | 0.0173 |
| **Hubei** | 0.000002 | 0.00002 | 0.00017 | 0.00042 | 0.0009 | 0.0039 | 0.0122 | 0.0265 | 0.0387 | 0.0157 |
| **Germany** | 0.000002 | 0.000008 | 0.00010 | 0.00037 | 0.0006 | 0.0022 | 0.0080 | 0.0204 | 0.0363 | 0.0110 |
| **NL** | 0.000002 | 0.000006 | 0.00008 | 0.00035 | 0.0005 | 0.0022 | 0.0080 | 0.0209 | 0.0373 | 0.0099 |

Own calculations based on the age-specific lower 95% IFR of Verity and colleagues[1] and abridged life tables of the UN World Population Prospects.[2]

**SI appendix 3**

**Age pattern of COVID-19 deaths**

Figure S2 shows the pattern over age of COVID-19 deaths using data provided by Dudel and colleagues.[3] Based on all available death age profiles, age standardized and normalized to sum to one, we calculate the average pattern over age that we use to split total deaths.

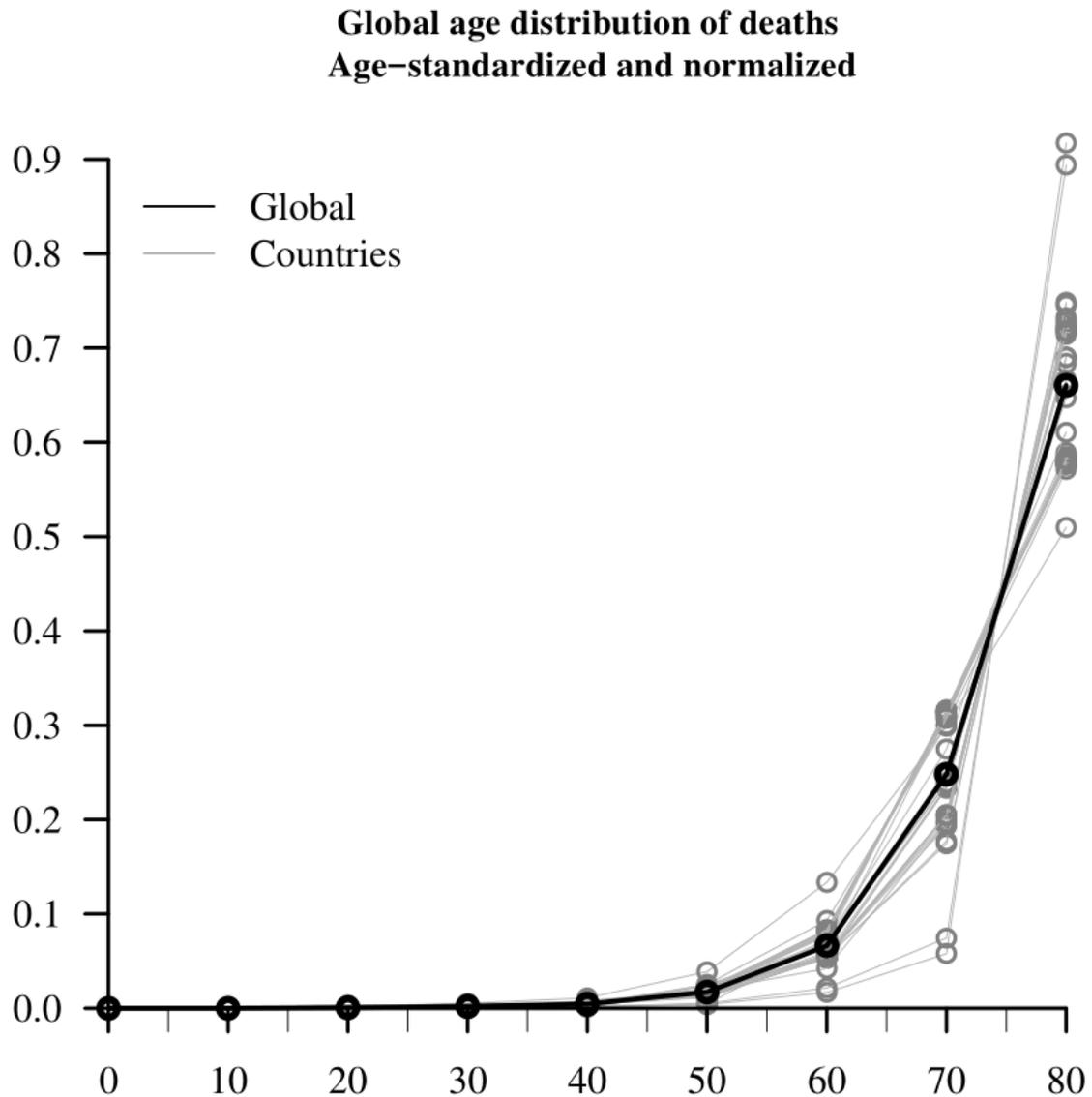

**Fig. S2.** Estimated global pattern over age of deaths attributable to COVID-19. Own calculations using data provided in Dudel and colleagues.[3]

**SI appendix 4**

**Estimation of COVID-19 infections using the unadjusted model**

Figures S3 and S4 illustrate the estimated numbers and population shares of COVID-19 infections based on scaling Chinese modal infection fatality rates via chronological age in the unadjusted model, from January 22 to May 13, 2020.

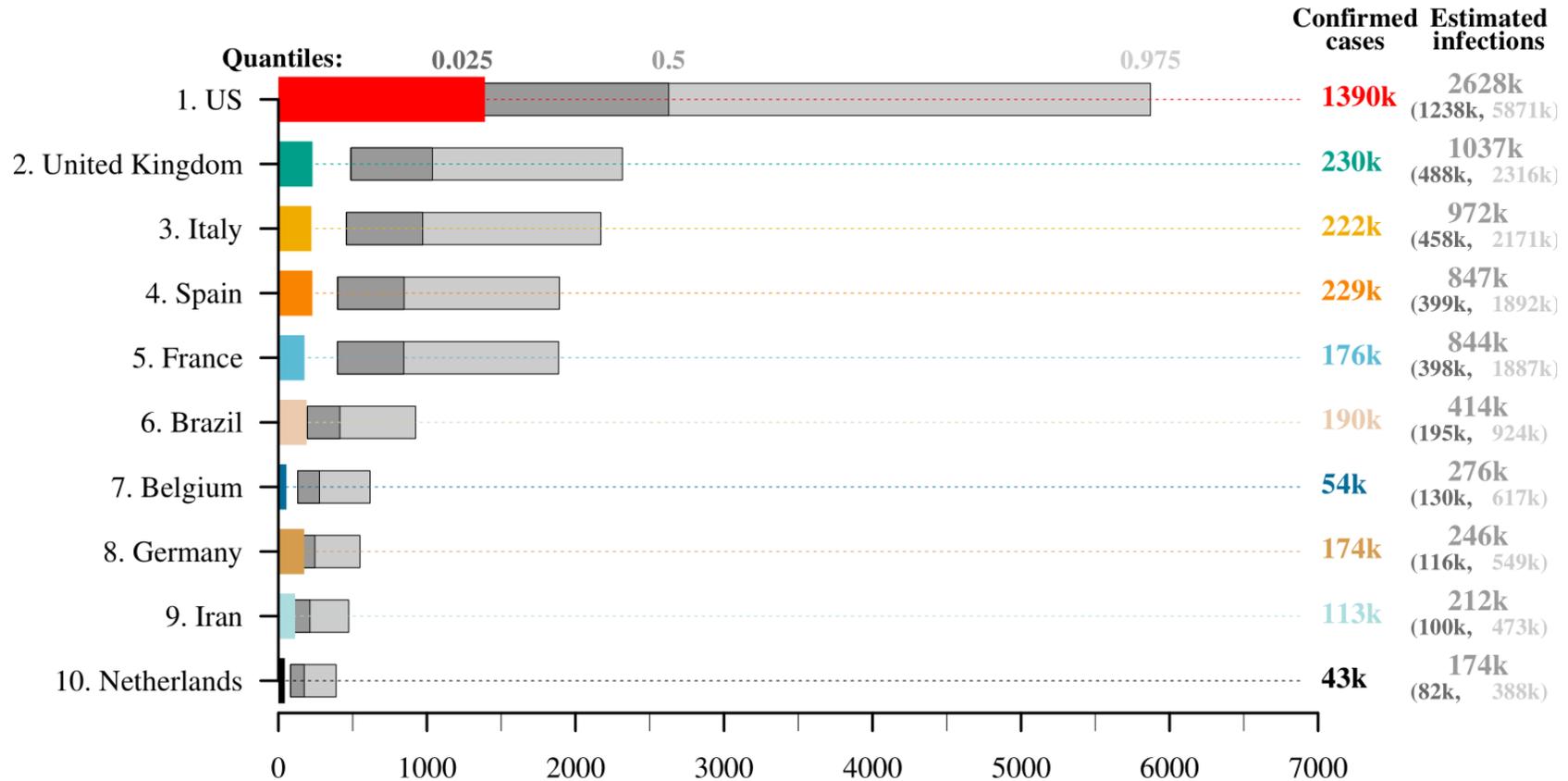

**Fig. S3.** Estimated numbers of COVID-19 infections based on unadjusted model using the Chinese modal, lower and upper 95% infection fatality rates, from January 22 to May 13, 2020. Shown are the 10 countries that have the largest numbers of reported deaths from COVID-19 as of May 13, 2020. Own calculations using data from Verity and colleagues,[1] UN World Population Prospects,[2] and Johns Hopkins University CSSE.[4]

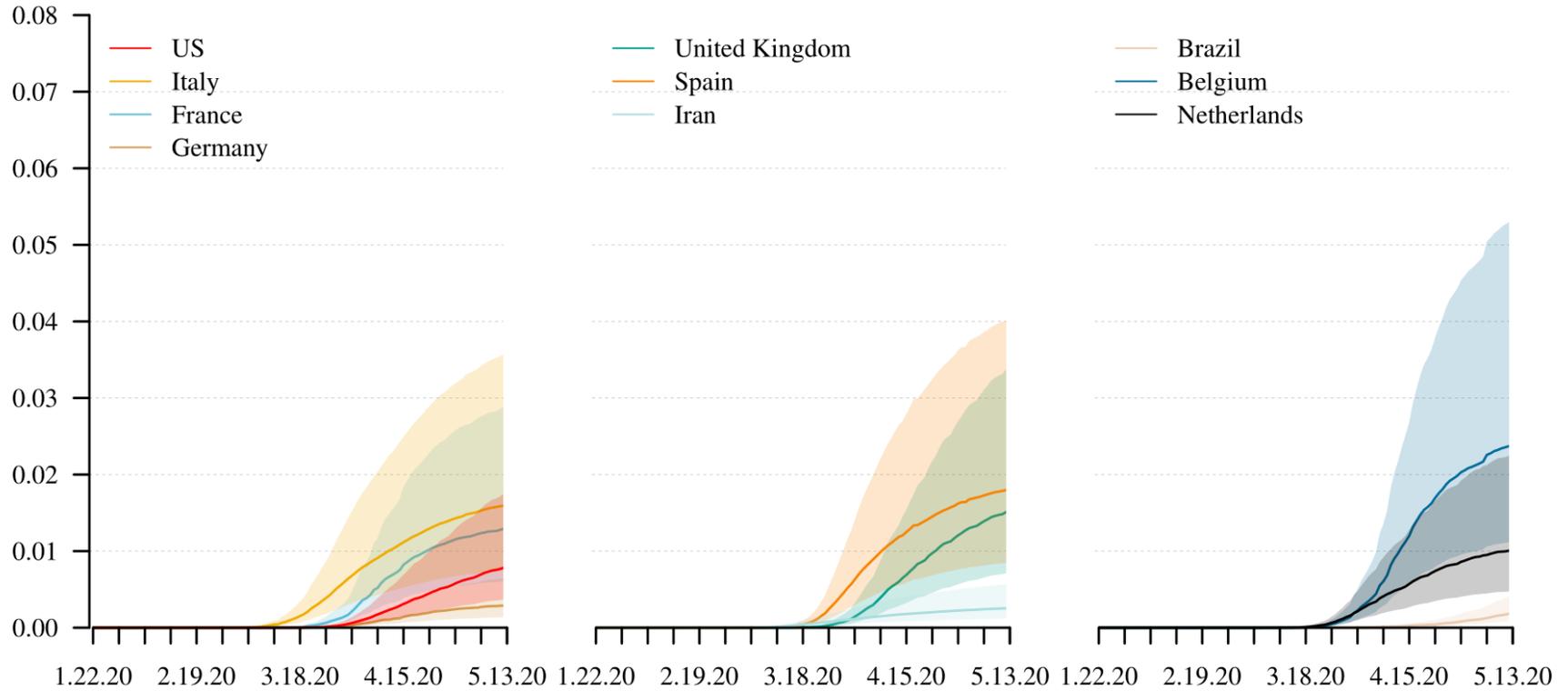

**Fig. S4.** Estimated prevalence of COVID-19 infections based on unadjusted model using the Chinese modal, lower and upper 95% infection fatality rates, from January 22 to May 13, 2020. Shown are the 10 countries that have the largest numbers of reported deaths from COVID-19 as of May 13, 2020. Own calculations using data from Verity and colleagues,[1] UN World Population Prospects,[2] and Johns Hopkins University CSSE.[4]

**SI appendix 5**

**Take French infection fatality rates as reference in adjusted model**

Figures S5 and S6 illustrate the estimated total numbers and population shares of COVID-19 infections based on the adjusted model that scales French infection fatality rates as reported by Salje and colleagues[5] via remaining life expectancy (thanatological age), from January 22 to May 13, 2020.

Across the 10 countries with the most COVID-19 deaths as of May 13, 2020, our central estimates suggest that the total number of infections is slightly more than six times higher than the number of confirmed cases. However, the level of uncertainty is high, as the lower bound of the 95% credible interval suggests that, on average, there are almost four times as many infections as confirmed cases, and the upper bound indicates that there are even more than 16 times as many infections as confirmed cases. Country-specific variation is high. Our modal estimates for Italy suggest that the total number of infections is approximately 2.1 million, or more than nine times higher than the number of country-specific confirmed cases. For the U.S., our modal estimate of 4.1 million infections is almost three times higher than the number of confirmed cases, and the upper bound of almost 13.7 million infections is almost 10 times higher than the number of confirmed cases. For Germany, where testing has been comparatively extensive, we estimate that the total number of infections is only 2.5 times higher (upper bound: 6.5 times higher) than the number of confirmed cases.

Based on the central estimates, as of May 13, 2020, we find that the prevalence of COVID-19 infections ranges from roughly 4% in Belgium and Spain; to between 3.4% and 2.7% in Italy, France, and the U.K.; to approximately 1.9% in the Netherlands; to 1.3% in the U.S. and to 0.5% or less in Germany, Iran, and Brazil. The uncertainty bounds are wide and range on average between 1.3% and 5.6%. The COVID-19 prevalence estimates are largest in Belgium, Spain, and Italy with 11.5%, 10.1%, and 8.7%, respectively.

We find that the total numbers and prevalence of infections are estimated to differ between the two sets of infection fatality rates as reported by Verity and colleagues[1] and Salje and colleagues[5]. The infection estimates appear to be larger based on infection fatality rates reported by Salje and colleagues.[5] Figure S7 compares the two sets of infection fatality rates. It shows that the infection fatality rates of Salje and colleagues[5] are lower for age groups 50-59, 60-69, and 70-79 and higher for age group 80+. Both sets of infection fatality rates are produced with a Bayesian model. They can be biased due to, e.g., model misspecification and misreported input data.

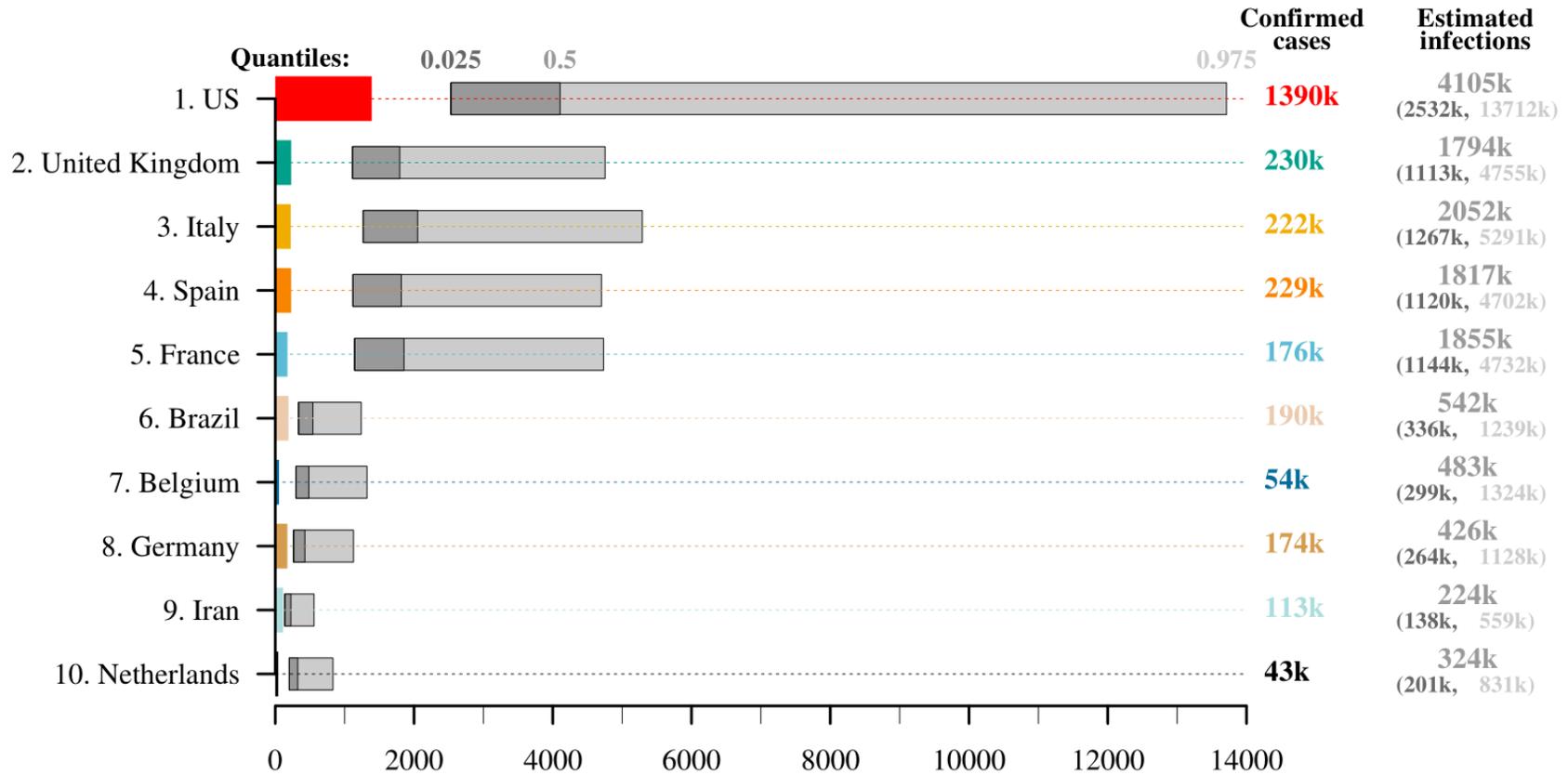

**Fig. S5.** Confirmed cases and estimated total number of COVID-19 infections based on adjusted model that scales French infection fatality rates from Salje and colleagues[5]. Shown are the 10 countries that have the largest numbers of reported deaths from COVID-19 as of May 13, 2020. Own calculations using data from Salje and colleagues,[5] UN World Population Prospects,[2] and JHU CSSE.[4]

**Fraction of people probably infected with COVID−19, January 22 − May 13, 2020**

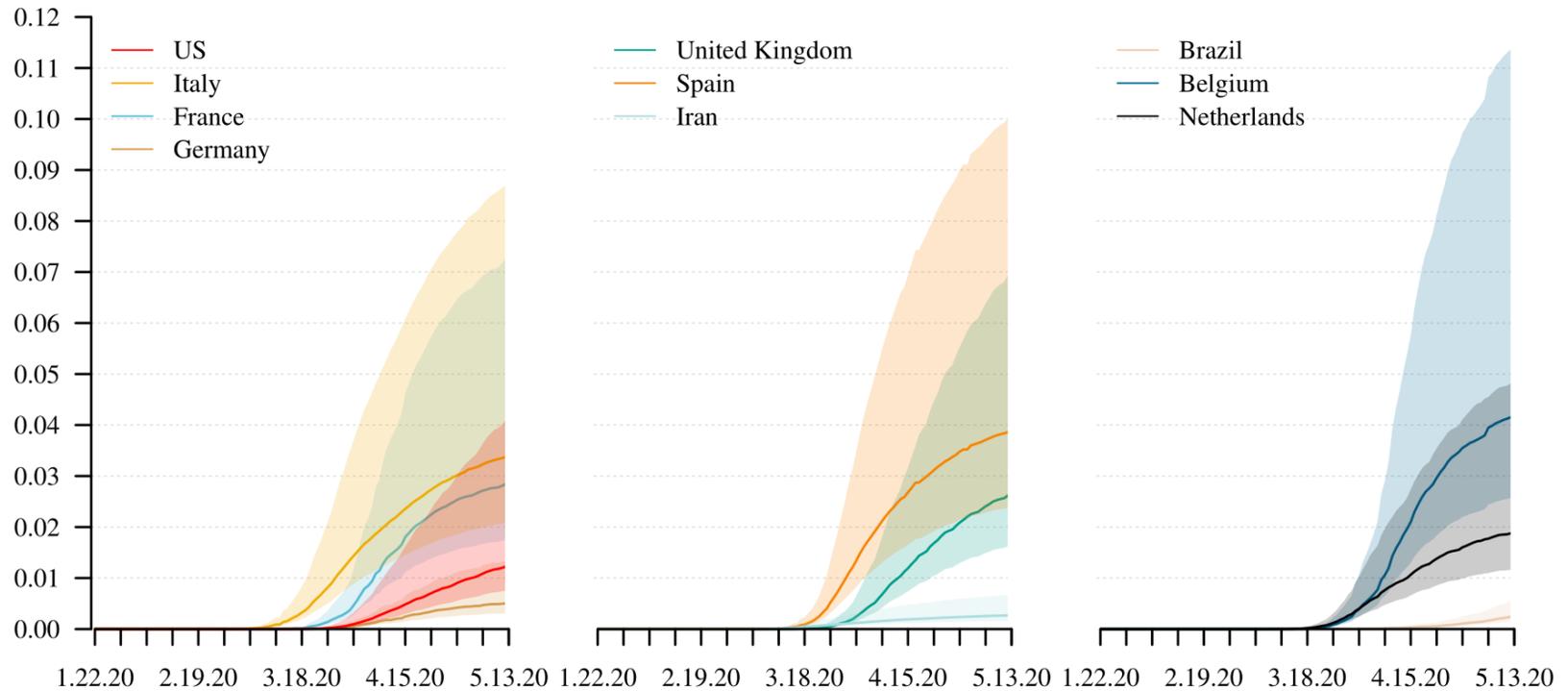

**Fig. S6.** Estimated population share of COVID-19 infections based on adjusted model that scales French modal, lower and upper 95% infection fatality rates from Salje and colleagues[5], from January 22 to May 13, 2020. Shown are the 10 countries that have the largest numbers of reported deaths from COVID-19 as of May 13, 2020. Own calculations using data from Salje and colleagues,[5] UN World Population Prospects,[2] and Johns Hopkins University CSSE.[4]

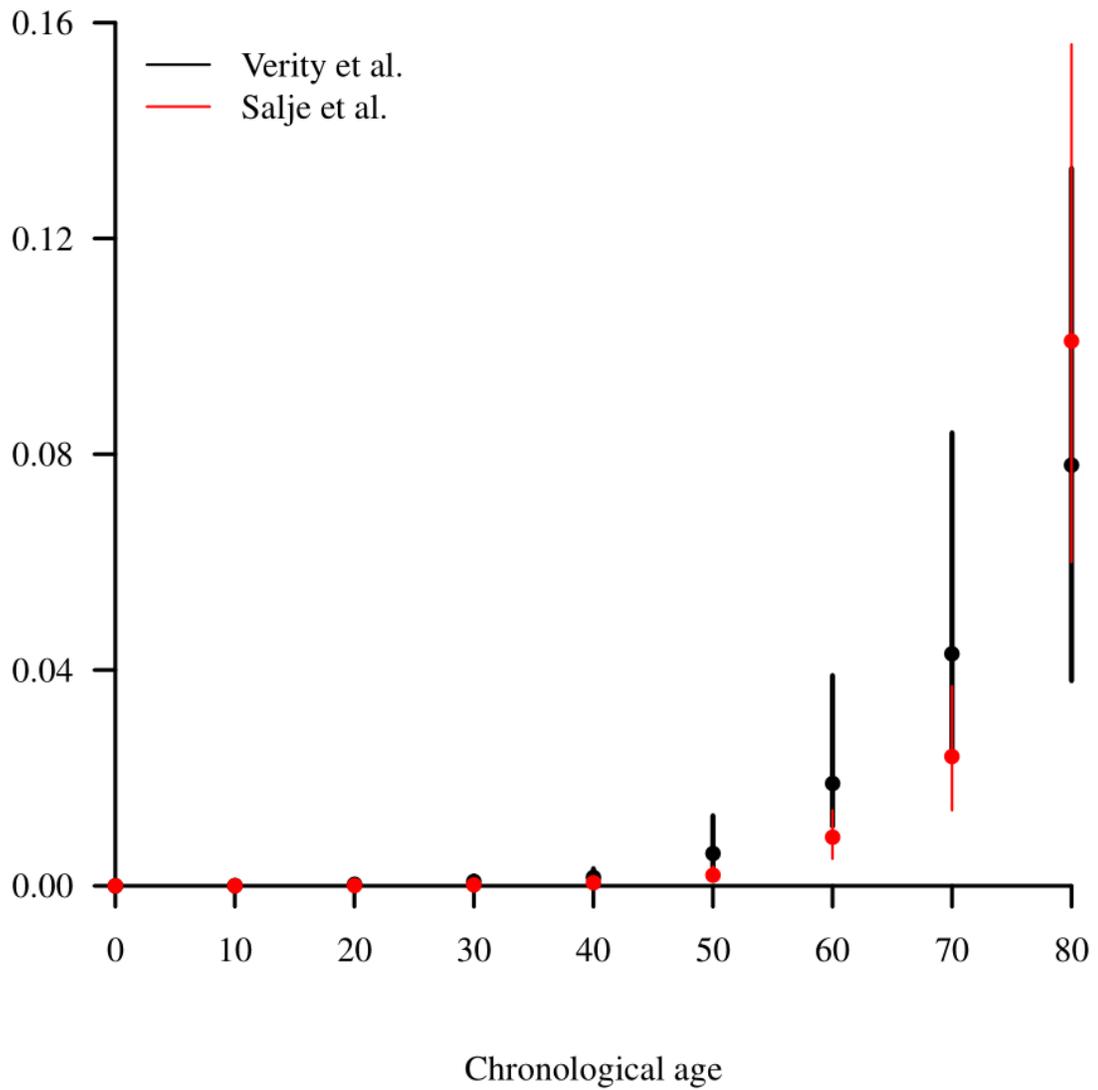

**Fig. S7.** Infection fatality rates (central estimate and 95% credible interval) reported by Verity and colleagues[1] (black) and by Salje and colleagues[5] (red) for 10-year age groups.

**SI appendix 6**

**Account for time to death when estimating COVID-19 infections**

Figure S8 compares the estimated numbers of COVID-19 infections with the numbers of confirmed cases as of April 26, 2020, in order to account for the average time to death of 18.5 days, as reported in Zhou and colleagues.[6] We use the adjusted model, scale the Chinese IFRs via remaining life expectancy (thanatological age), and take the latest deaths from COVID-19 as of May 13, 2020.

Across the 10 countries with the most COVID-19 deaths as of May 13, 2020, our central estimates suggest that the total number of infections is approximately six times higher than the number of confirmed cases. However, the level of uncertainty is high, as the lower bound of the 95% credible interval suggests that, on average, there are three times as many infections as confirmed cases, and the upper bound indicates that there are even more than 14 times as many infections as confirmed cases. Country-specific variation is high. Our modal estimates for Italy suggest that the total number of infections is approximately 1.4 million, or more than seven times higher than the number of country-specific confirmed cases. For the U.S., our modal estimate of 3.1 million infections is more than three times higher than the number of confirmed cases, and the upper bound of 6.7 million infections is more than seven times higher than the number of confirmed cases. For Germany, where testing has been comparatively extensive, we estimate that the total number of infections is only two times higher (upper bound: close to five times higher) than the number of confirmed cases.

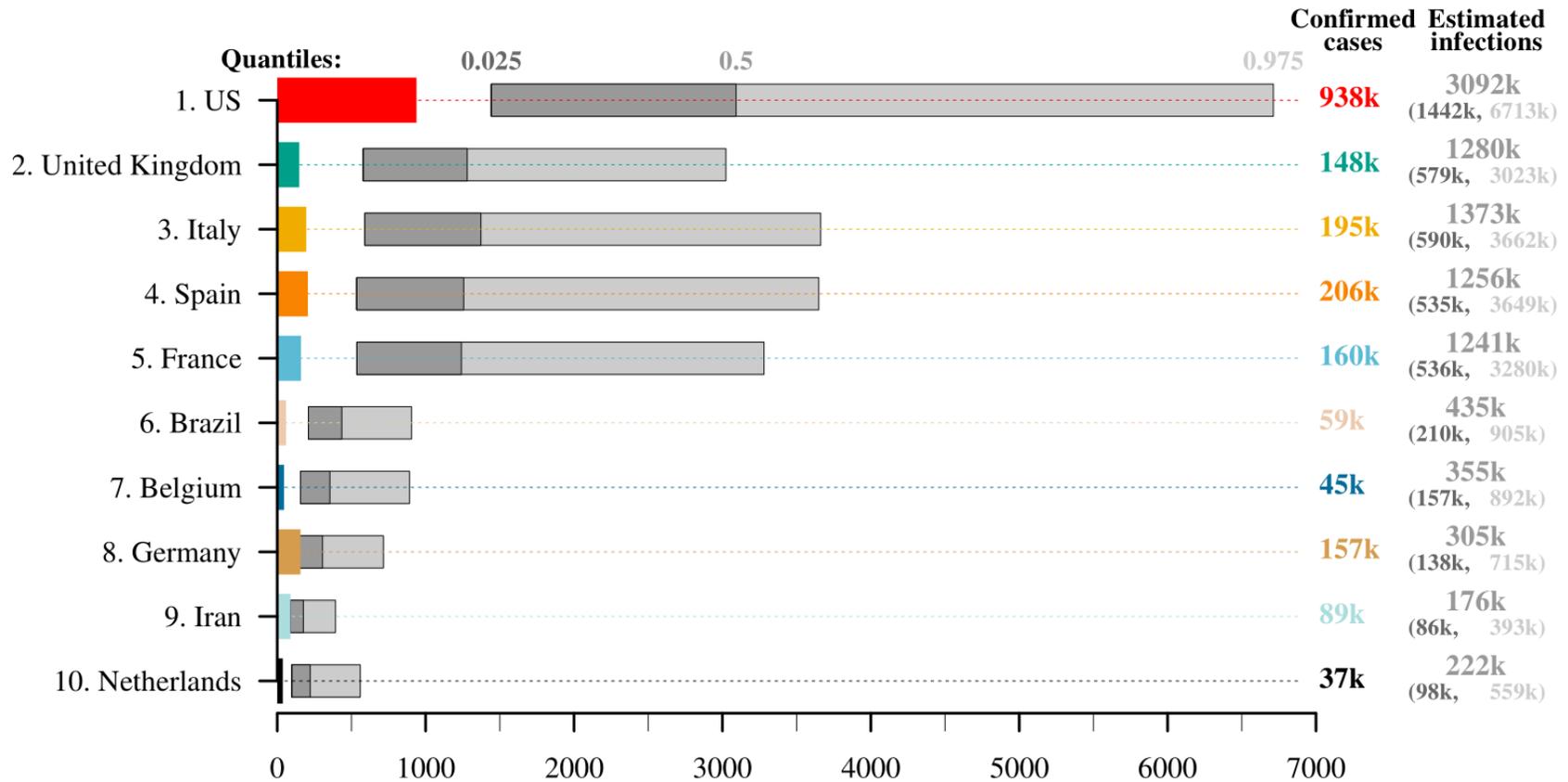

**Confirmed cases of April 26, 2020, vs estimated infections of May 13, 2020, in thousand**

**Fig. S8.** Confirmed cases, as of April 26, 2020, and estimated total number of COVID-19 infections, as of May 13, 2020, to account for average time to death: 18 days. Estimations are based on adjusted model that uses COVID-19 deaths as of May 13, 2020, and scaled Chinese infection fatality rates. Data are shown for the 10 countries that had the largest numbers of reported deaths from COVID-19 as of May 13, 2020. Own calculations using data from Verity and colleagues[1], UN World Population Prospects,[2] and JHU CSSE.[4]

**SI References**